\documentclass[12pt]{iopart}
\usepackage{iopams}  
\usepackage{graphicx}
\begin{document}

\title{Optical tweezers: wideband microrheology}

\author{Daryl Preece $^{1}$, Rebecca~Warren$^{2}$, Manlio Tassieri$^{2}$,
R.M.L.~Evans$^{3}$, Graham~M.~Gibson$^{1}$, Miles~J.~Padgett$^{1}$,
Jonathan~M.~Cooper$^{2}$}
\address{$^{1}$Department of Physics and Astronomy, SUPA, University of Glasgow,
G12 8QQ, U.K.}
\address{$^{2}$Department of Electronics and Electrical Engineering, University
of Glasgow, G12 8LT, U.K.}
\address{$^{3}$School of Physics and Astronomy, University of Leeds, LS2 9JT,
U.K.}
\ead{m.tassieri@elec.gla.ac.uk}
\ead{d.preece@physics.gla.ac.uk}

\date{xx xxxx, 2010}

\begin{abstract}
Microrheology is a branch of rheology having the same principles as conventional
bulk rheology, but working on micron length scales and $\mu l$ volumes. 

Optical tweezers have been successfully used with Newtonian fluids for
rheological purposes such as determining fluid viscosity. Conversely,
when optical tweezers are used to measure the viscoelastic properties of
complex fluids the results are either limited to the material's high-frequency
response, discarding important information related to the low-frequency
behaviour, or they are supplemented by low-frequency measurements performed with
different techniques, often without presenting an overlapping region of clear
agreement between the sets of results.
%supplimented
We present a simple experimental procedure to perform microrheological
measurements over the widest frequency range possible with optical tweezers. A generalised Langevin equation is used to
relate  the frequency-dependent moduli of the complex fluid to the
time-dependent trajectory of a probe particle as it flips between two optical
traps that alternately switch on and off.
%edited above
\end{abstract}

%\pacs{83.60.Bc, 66.20.-d, 83.50.-v, 83.85.Ei}

\maketitle
\section{Introduction}

Optical tweezers have become an increasingly important
tool for the manipulation of micron-scale objects. Taking advantage
of the optical gradient force \cite{Ashkin1992} they have been used both
as a force sensor \cite{Molloy1995, Block1989} and as an force
actuator \cite{Guck:2001, Wang}. Though the single
beam gradient trap has been around since Ashkin \cite{Ashkin86}, modern
holographic
optical tweezers, which use spatial light modulators to create multiple
optical traps, are capable of moving multiple micron-sized objects
in three dimensions \cite{Dufresne2001b}. Recent advances in SLM technology mean that optical
tweezers are now capable of updating trap positions at hundreds of
frames per second with low latency ($\approx$5ms) \cite{Preece}. This can only
serve to broaden the already extensive range of applications for which
optical tweezers have been used, such as cell biology \cite{Svoboda},
investigations into DNA \cite{Wang}, and driving micro-machines
\cite{Friese2001}.

Along with biological and physical experiments optical tweezers have
also proved useful for microrheological experiments \cite{Bishop2004}.
Microrheology is concerned with the linear viscoelastic response of materials
at microscopic length scales. Such information is invaluable for the
investigation
of unknown biological processes as well as for increased understanding
of the basic physics of fluids. The linear viscoelastic properties
of a fluid can be represented by the frequency-dependent dynamic bulk
modulus $G^{*}(\omega)$. The dynamic bulk modulus encompasses information
about both the elastic and viscous properties of the fluid. It is
expressed in the form $G^{*}(\omega)=G'(\omega)+iG''(\omega)$
\cite{Ferry},
where $G'(\omega)$ is the frequency-dependent elastic modulus
and $G''(\omega)$ is the frequency-dependent viscous modulus \cite{Tassieri}. 

As is well known~\cite{Neuman,Di2008,Berg}, the viscosity of
the medium in which a trapped particle is immersed is a key factor
affecting the statistical distribution of the particle's time-dependent
position. Hence the viscosity is easily determined from measurements of that
distribution. Furthermore, surface effects such as Fax\'en's correction to the
viscosity of a fluid close to an
infinite surface (\cite{Di2008,Berg,Yao}) have also been widely studied in Newtonian
fluids.
However when microrheological measurements of \emph{viscoelastic} substances
have been made, results are either limited the high end of the frequency
response \cite{Starrs,Atakhorrami,Nijenhuis} (omitting the low frequencies entirely)
or account is taken of the low frequency response only by the
use of complementary techniques such as rotational rheometry \cite{Pesce2009}
or passive video particle tracking microrheology \cite{Marija}. In
both cases this has been done without showing an overlapping region, thus
leaving a significant information gap.

The aim of this paper is to present a self-consistent procedure for measuring
the linear viscoelastic properties of materials across the widest frequency
range achievable with optical
tweezers. In particular, the procedure consists of two steps: (I) measuring the
thermal fluctuations of a trapped bead for a sufficiently long time; (II)
measuring the transient displacement of a bead flipping between two optical
traps (spaced at fixed distance $D_0$) that alternately switch on/off
at sufficiently low frequency.
The analysis of the first step (I) provides: (a) the traps stiffness
($\kappa_{i}$, $i=1,2$) --- note that this has the added advantage of making the
method self-calibrated --- and (b) the high frequency viscoelastic
properties of the material, to high accuracy. 
The second step (II) has the potential to provide information about the
material's
viscoelastic properties over a very wide frequency range, which is only limited
(at the top end) by the acquisition rate of the bead position ($\sim$kH) and
(at the bottom end) by the duration of the experiment. However, because of the
finite time required by the equipment to switch on/off (i.e.~tens of
milliseconds), the material's high-frequency response can not be fully
determined by this step. The full viscoelastic spectrum is thus resolved by combining
the results obtained from steps (I) and (II).

\section{Analytical model}

To understand the basis of the procedure, we consider
the time-dependent position $\vec{r}(t)$ of a bead trapped by a stationary
harmonic
potential of force-constant $\kappa_{i}$. 
Throughout the first step (I) of the procedure, the bead is always found close
to the centre of the single trap which is switched on; it makes only small
deviations $\vec{r}(t)$ (of magnitude set by the thermal energy) away from the
centre of the trap. During step (II) of the procedure, the detours are
considerably larger, of a magnitude set by the separation $D_0$ of the traps.
Let us define $t_{\rm II}$ to be the time at which step (II) commences, i.e.~the
time at which the traps are first switched. Subsequently, each trap remains on
for a duration $P$ before it is switched off and the other trap switched on.
Hence the total period of the repeated sequence is $2P$. At the instant
immediately after the traps are switched (i.e.~at $t=t_{\rm II}+nP$ where
\mbox{$n=0,1\ldots N$}), the bead is typically positioned
at a distance $\left|\vec{r}(t)\right|\approx D_0$  from the centre of the
currently active trap (i.e.~close to the centre of the trap that has just
switched off). Note that coordinates are re-defined so that the bead's
displacement $\vec{r}(t)$ is always measured with respect to the centre of
whichever trap is currently switched on.
The bead's position in three dimensions can be modelled by a generalized
Langevin equation,
\begin{equation}
\label{Langevin}
m\vec{a}(t)=\vec{f}_{R}(t)-\int^{t}_{t_0}\zeta(t-\tau){\vec{v}}
(\tau)d\tau-\kappa_{i} \vec{r}(t),
\end{equation}
where $m$ is the mass of the particle, $\vec{a}(t)$ is its acceleration,
$\vec{v}(t)$ its velocity and $\vec{f}_{R}(t)$ is the usual Gaussian
white noise term, modelling stochastic thermal forces acting on the particle.
The integral, which incorporates a generalized time-dependent memory function
$\zeta(t)$, represents viscoelastic drag from the fluid.

Note that the present method does not require the two traps to be equal. They
can be independently, but not simultaneously, calibrated during step I by
appealing to the Principle of Equipartition of Energy,
\begin{equation}
\label{Equipartition}
	\frac{3}{2}k_{B}T=\frac{1}{2}\kappa_{i} \langle r^2 \rangle,
\end{equation}
where $k_{B}$ is Boltzmann's constant, $T$ is absolute temperature and $\langle
r^2 \rangle$ is the time-independent variance of the particle's displacement
from
the trap's centre. Despite the variety of established methods for
determining the stiffness of an optical trap (e.g.~using the power spectrum or
the
drag force \cite{Berg,Neuman}), the Equipartition method is the only one
independent of the viscoelastic properties of the material under investigation
and is thus essential to the calibration of a rheological measurement.

We now examine how Eq.~(\ref{Langevin}) evolves in the two steps (I) and (II).
During step (I), the thermal fluctuations of the trapped bead are
investigated to determine the high-frequency viscoelastic properties of the
material through analysis of the time dependence of the normalised position
autocorrelation function $A(\tau)$ (``NPAF"):
\begin{equation}
\label{NPAFt}
	A(\tau)=\frac{\left\langle
\vec{r}(t_0)\vec{r}(t_0+\tau)\right\rangle_{t_0}}{\left\langle
r(t_0)^2\right\rangle_{t_0}},
\end{equation}
which, by time-translation invariance, is a function only of the time interval
$\tau$. Here, $\left\langle r(t_0)^2\right\rangle_{t_0}$ is the time-independent
variance and the brackets $\left\langle...\right\rangle_{t_0}$ denote an average
over all initial times $t_0$.

Multiplying both the sides of Eq.~(\ref{Langevin}) by
$\vec{r}(t_0)$ and averaging over $t_0$, one obtains
\begin{equation}
\label{NPAFL}
	\tilde{A}(s)=\left(s+\frac{\kappa_{i}}{ms+\zeta(s)}\right)^{-1},
\end{equation}
where $\tilde{A}(s)$ is the Laplace transform of $A(\tau)$, $s$ is the Laplace
frequency and it has been assumed that both 
the quantities $\left\langle \vec{r}(t_0)\vec{v}(t_0)\right\rangle_{t_0}$ and
$\left\langle\vec{r}(t_0)\vec{f}_{R}(t_0+\tau)\right\rangle_{t_0}$ vanish for
all $\tau$.

We note that, for $ms\ll\tilde{\zeta}(s)$ (which is a good approximation up
to MHz frequencies for micron sized beads with density of order $1 g/cm^3$ in
aqueous suspension) and for a Newtonian fluid (i.e. a liquid with
time-independent viscosity
$\eta$, for which $\zeta=6\pi \eta a$), Eq.~(\ref{NPAFL}) recovers the
well-known result for a massless particle harmonically trapped in a Newtonian
fluid,
\begin{equation}
\label{NPAFN}
	A(\tau)\to\exp\left(-\Gamma_{i}\tau\right),
\end{equation}
where $\Gamma_{i}=\kappa_{i}/6 \pi a \eta$ is the characteristic relaxation rate
of the system and can be used to determine $\eta$ once both $\kappa_{i}$ and $a$
are known.
% changed eta to teta tau
In the general case of non-Newtonian fluids (i.e.~materials with time-dependent
viscosity $\eta (t)$) we follow Mason and Weitz \cite{Mason} in assuming that
the bulk Laplace-frequency-dependent viscosity of the fluid $\tilde{\eta}(s)$ is
proportional to the microscopic memory function 
$\tilde{\zeta}(s)=6\pi a
\tilde{\eta}(s)$, so Eq.~(\ref{NPAFL}) can be written as
\begin{equation}
\label{eta}
	\tilde{\eta}(s)=\frac{\kappa_{i}}{6\pi
a}\left[\frac{\tilde{A}(s)}{\left(1-s\tilde{A}(s)\right)}-\frac{ms}{\kappa_{i}}
\right].
\end{equation}
Moreover, given that 
$G^*(\omega) \equiv \left. s\tilde{\eta}(s) \right|_{s=i\omega}$,
the complex viscoelastic modulus $G^*(\omega)$ can be expressed directly in
terms of the time-dependent NPAF,
\begin{equation}
\label{G*NPAF}
	G^*(\omega)=\frac{\kappa_{i}}{6\pi
a}\left[\frac{i\omega\hat{A}(\omega)}{\left(1-i\omega\hat{A}(\omega)\right)}
+\frac{m\omega^{2}}{\kappa_{i}}\right],
\end{equation}
where $\hat{A}(\omega)$ is the Fourier transform of $A(\tau)$ and, as mentioned
before, the inertia term ($m\omega^{2}$) can be neglected for frequencies
$\omega\ll$MHz.

It is interesting to highlight that Eq.~(\ref{G*NPAF}) is actually equivalent
to Eq.~(6) in Ref.~\cite{Tassieri}. Indeed, $A(\tau)$ is directly related to the
normalised mean-square displacement 
$\left\langle \Delta r^{2}(\tau)\right\rangle / 2\left\langle r^2\right\rangle$ 
introduced in Ref.~\cite{Tassieri} as follow:
\begin{eqnarray}
\label{MSD-A}
  \frac{\left\langle \Delta r^{2}(\tau)\right\rangle}{2\left\langle
  r^2\right\rangle}=\frac{\left\langle r^{2}(\tau)\right\rangle+\left\langle
  r^{2}(t_0)\right\rangle-2\left\langle
  \vec{r}(t_0)\vec{r}(\tau)\right\rangle}{2\left\langle r^2\right\rangle}=
  \nonumber \\ 
  =1-A(\tau).~~~~~~~~~~~~~~~~~~~~~~~~~~~~~~~~~~~~~~~~~~~~~~~~~
%check eqn
\end{eqnarray}
By performing the Fourier transform of Eq.~(\ref{MSD-A}) one obtains
the relation
\begin{equation}
\label{MSD-A2}
  \frac{\left\langle \Delta \widehat{r^2}(\omega)\right\rangle}{2\left\langle
  r^2\right\rangle}=\frac{1}{i\omega}-\hat{A}(\omega).
\end{equation}
Note that the quantity $A(\tau)$ has the added advantage of having a
well-controlled
Fourier transform unlike the $MSD(\tau)$ as discussed below.

The second step (II) of the procedure consists of analysing the bead's transient
displacements as it moves between two traps with separation $D_0$ that swap
their on/off state at times $t=t_0+nP$. Note that the duration $P$ must exceed
all of the material's characteristic relaxation times. We define the normalized
mean position of the particle as
$D(t)=\left|\left\langle\vec{r}(t)\right\rangle\right|/D_{0}$, where the
brackets $\left\langle ...\right\rangle$ denote the average over several
independent measurements, but not over absolute time, since
time-translation invariance is broken by the periodic switching.
In this case, Equation~(\ref{Langevin}) yields, in the Laplace form, an
identical expression to Eq.~(\ref{NPAFL}) with $\tilde{A}(s)$ replaced by
$\tilde{D}(s)$, the Laplace transform of $D(t)$. Thus, as before,
the complex modulus can be expressed directly in terms of the bead position:
\begin{equation}
\label{G*r}
	G^*(\omega)=\frac{1}{12\pi
a}\sum_{i=1,2}\left[\frac{\kappa_{i}i\omega\hat{D}(\omega)}{\left(1-i\omega\hat{
D}(\omega)\right)}+m\omega^{2}\right],
\end{equation}
where $\hat{D}(\omega)$ is the Fourier transform of $D(t)$ and the sum takes
account of the linearity of the measurements performed with both traps.

Note that both functions $A(\tau)$ and $D(t)$ are expected to have the limits
$A(0)=D(0)=1$ and $A(\infty)=D(\infty)=0$. This turns out to be very useful
when 
applying the Fourier transform during data processing, as shown below.

In principle, Eqs.~(\ref{G*NPAF}) and (\ref{G*r}) are two simple
expressions relating the material's complex modulus $G^*(\omega)$ to the
observed
time-dependent bead trajectory $\vec{r}(t)$ \textit{via} the Fourier transform
of either $\vec{r}(t)$ itself (in Eq.~(\ref{G*r})) or the related NPAF (in
Eq.~(\ref{G*NPAF})).
In practice, the evaluation of these Fourier transforms, given only a finite set
of data points over a finite time domain, is non-trivial since interpolation and
extrapolation from those data can yield serious artefacts if handled carelessly.

The $N$ experimental data points $(t_k,A(t_k))$ or
$(t_k,D(t_k))$, where \mbox{$k=1\ldots N$},  extend over a
finite range of times $t$, exist only for positive $t$ and {\em need not} be
equally spaced.
In order to express the Fourier transforms in Eqs.~(\ref{G*NPAF}) and
(\ref{G*r}) in terms of these data,
we adopt the analytical method introduced in Ref.~\cite{Evans}. In particular,
we refer to Eq.~(10) of Ref.~\cite{Evans} which is equally applicable to find
the Fourier transform $\hat{g}(\omega)$ of any time-dependent quantity $g(t)$
sampled at a finite set of data points $(t_k,g_k)$, giving
\begin{eqnarray}
\label{Fg}
  -\omega^2 \hat{g}\left(\omega\right) =  i\omega g(0) + \left( 1-e^{-i\omega
t_1}\right) \frac{\left(g_1-g(0)\right)}{t_1} \nonumber \\
  + \dot{g}_\infty e^{-i\omega t_N} + \sum_{k=2}^N \left(
\frac{g_k-g_{k-1}}{t_k-t_{k-1}} \right) \left( e^{-i\omega t_{k-1}}-e^{-i\omega
t_k} \right),
\end{eqnarray}
where $\dot{g }_\infty$ is the gradient of $g (t)$ extrapolated to infinite
time. Also $g (0)$ is the value of $g(t)$ extrapolated to $t=0^+$.
Identical formulas can be written for both $\hat{A}(\omega)$ and
$\hat{D}(\omega)$, with $g$ replaced by $A$ and $D$ respectively. 
It is a strength of our new procedure that both of the extrapolated quantities 
$\dot{g}_\infty$ and $g(0^+)$ (which assumed the role of fitting parameters in
Ref.~\cite{Evans}) in this case assume known values ($\dot{g}_\infty=0$ and
$g(0^+)=1$) given by the limits of the functions $A(\tau)$ and $D(t)$. Like the
method presented in Ref.~\cite{Evans}, the present procedure also
has the advantage of removing the need for Laplace/inverse-Laplace
transformations of experimental data \cite{Mason2}.

\section{Experimental Setup}

\begin{figure}
  \begin{centering}
  \includegraphics{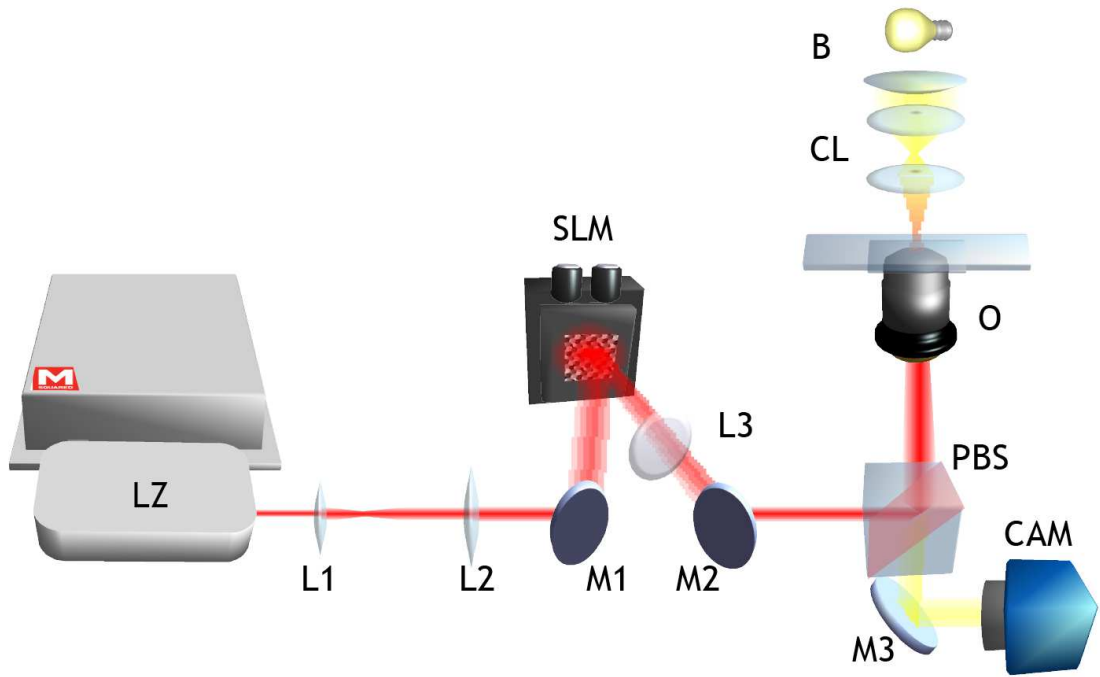}
  \par
  \end{centering}
  \caption{\label{fig:rheology-set-up}Experimental setup. From left to right.
  (LZ) Ti:sapphire laser system, (L1 and L2) Beam Telescope, (M1 and
  M2) Folding Mirrors, (SLM)Boulder Fast SLM,(L3), (PBS) Polarising beam
  splitter cube, (M3) Mirror, (CAM) Prosilica Fast camera, (O) Objective
  lens, (CL)Condensing optics, (B) 250W halogen bulb.}
\end{figure}

We have validated the experimental procedure as described by Eqs.~(\ref{G*NPAF}) and (\ref{G*r}), \textit{via}
Eq.~(\ref{Fg}),  by measuring both the viscosity of water and the viscoelastic
properties of water-based solutions of polyacrylamide (PAM, flexible
polyelectrolytes, $M_w=5$~---~$6\times10^6$~g/mol, Polysciences Inc.) using
optical tweezers as described below and shown schematically in
Fig.~\ref{fig:rheology-set-up}.

Trapping is achieved using a CW Ti:sapphire laser system (M Squared,
SolsTiS) which provides up to 1 W at 830~nm. Holographic optical
traps are created via the use of a spatial light modulator (Boulder
XY series) \cite{Preece} in the Fourier plane of the optical traps.
The tweezers are based around an inverted microscope, where the same
objective lens, 100$\times$ 1.3NA, (Zeiss, Plan-Neofluor) is used
both to focus the trapping beam and to image the resulting motion
of the particles. Samples are mounted in a motorized microscope stage
(ASI, MS-2000). Particles are imaged using bright-field illumination.
We use a Prosilica GC640M camera to view the trapped particle and
our own suite of camera analysis software written in LabVIEW (available from \cite{optics}) to measure
the position of the trapped particle in real time at up to 2~kHz frame rate \cite{Gibson}.

\section{Results}

\begin{figure}
	\centering
		{\includegraphics[width=10cm,height=10cm,keepaspectratio]{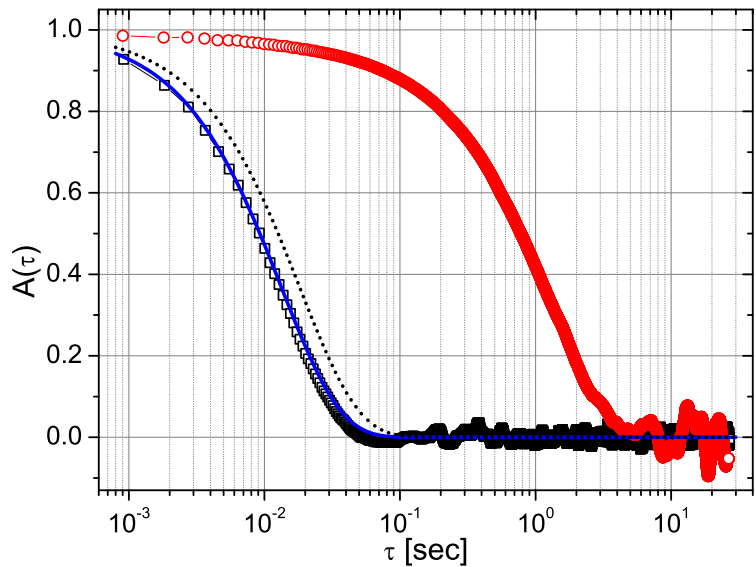}}
	\caption{\label{NPAF} The normalised position autocorrelation function
\textit{vs.} lag-time of a $5 \mu m$ diameter bead (squares) in water (with
$\kappa=2.7 \mu N/m$) and (circles) in a water-based solution of PAM at
concentrations of $1$ \% $w/w$ (with $\kappa=2.2 \mu N/m$). The continuous
and dotted lines represent Eq.~(\ref{NPAFN}) for a $5 \mu m$ diameter bead in
water at $T=25^o C$ with $\kappa=2.7 \mu N/m$ and $\kappa=2.2 \mu N/m$,
respectively.}	
\end{figure}

\begin{figure}
	\centering
		{\includegraphics[width=10cm,height=10cm,keepaspectratio]{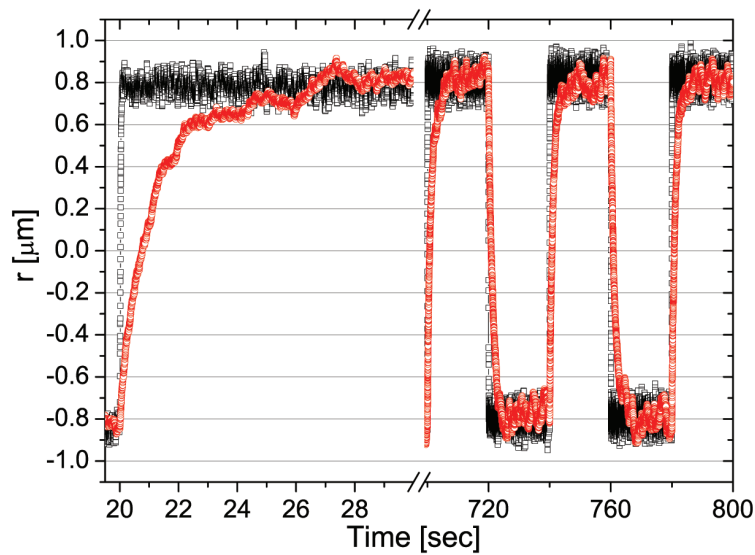}}
	\caption{\label{Steps} The trajectory of a $5 \mu m$ diameter bead
flipping between two optical traps $\kappa_{1}$ (bottom) and $\kappa_{2}$ (top)
repeatedly switching after a duration $P=20$~s. The bead is suspended in
(squares) water (with $\kappa_{1}=2.7$ and $\kappa_{2}=2.5 \mu N/m$) and
(circles) a water-based solution of PAM at concentrations of $1$ \% $w/w$ (with
$\kappa_{1}=2.1$ and $\kappa_{2}=2.2 \mu N/m$).}	
\end{figure}

\begin{figure}
	\centering
		{\includegraphics[width=10cm,height=10cm,keepaspectratio]{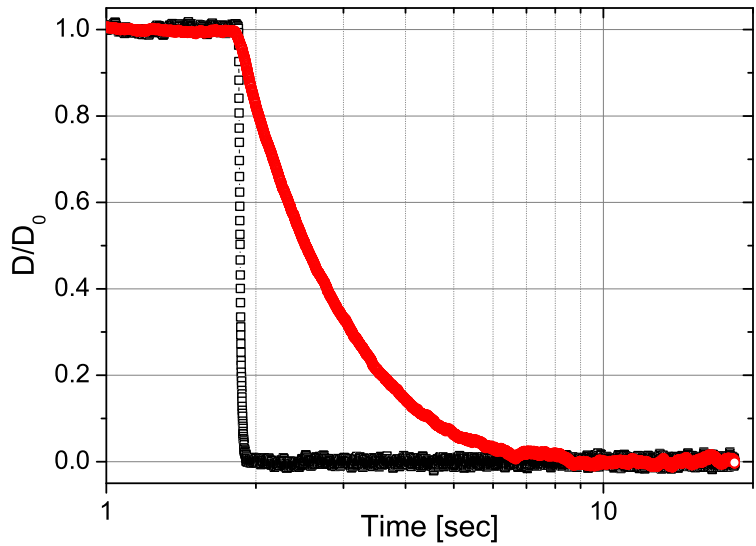}}
	\caption{\label{StepN} The normilized mean position of all step-down
data shown in Fig.~\ref{Steps}; i.e., when simultaneously trap 2 (top) switches
\textit{off} and trap 1 (bottom) switches \textit{on}.}	
\end{figure}

The normalized position autocorrelation functions
$A(\tau)$, measured from the stochastic fluctuations of beads optically trapped
in two different fluids are shown in Fig.~\ref{NPAF}, together with the function
predicted for a simple Newtonian fluid. In the Newtonian case, it is expected
that $A(\tau)$
decays as a single exponential with a characteristic relaxation rate related to
the trap strength, bead size and fluid viscosity i.e.~Eq.~(\ref{NPAFN}). The
agreement between the data and prediction for water is good. On the
other hand, in the case of a non-Newtonian fluid, where the viscosity is 
time-dependent, it is not guaranteed that Eq.~(\ref{NPAFL}) could
be resolved (i.e.~inverse-Laplace transformed) into a simple form like
Eq.~(\ref{NPAFN}). However, this is no hindrance since the viscoelastic moduli
will be found via the analysis of the normalized position autocorrelation
function.
%(how ''good`` is the data.)
In Fig.~\ref{Steps} we compare the impulse response (i.e.~step II of the
procedure)
%(changed to impulse response.)
of a $5 \mu m$ diameter bead suspended in water and in an aqueous solution of
PAM at 1\% w/w (a
non-Newtonian fluid), with a duration between flips of $P=20$~s and a trap
centre-to-centre separation of $D_{0}=1.6 \mu m$, giving $D_{0}/a=0.64$.
In order to guarantee the linearity of the confining forces exerted by the two
optical traps, the distance between them was
always chosen to be no more than 80\% of the bead radius,
$D_{0} \leq 0.8 a$ \cite{Ashkin1992}. Although Brownian statistical fluctuations
appear in the figure, the difference between the viscoelastic natures of the two
fluids is clear.
Indeed, while a bead suspended in water flips from one trap to the other
almost instantaneously, the same bead in the PAM
solution takes much longer (a few seconds) to flip. In order both to evaluate
$D(t)=\left|\left\langle\vec{r}(t)\right\rangle\right|/D_{0}$ and to reduce
noise caused by Brownian fluctuations, the transient measurements were averaged
over
twenty flips, with the resulting curves shown in Figure~\ref{StepN}.
Experimentally, the switching process of the two traps is controlled by means of
a spatial light modulator (SLM) which alternately creates an optical trap in one
of two positions. It is important at this point to note that there is a short
but finite time for which both traps exist simultaneously \cite{Preece}, due to the finite
time required by the SLM's 
display to update the holographic pattern.
This makes the switching process not exactly
binary. However, this will of course only affect the high-frequency results
obtained during
this step which will ultimately be neglected in favour of the high-frequency
response from step (I).

Wideband microrheological measurement are obtained from the optical tweezers by
combining the frequency responses obtained from both steps (I) and (II) of the
procedure.
In particular, the material's high-frequency response is determined by applying
Eq.~(\ref{G*NPAF}) (\textit{via} Eq.~(\ref{Fg}) with $A_k$ replacing $g_k$) to
the $A(\tau)$ measurements (in which low-frequency information tends to be very
noisy); whereas, the low-frequency response is resolved by
applying Eq.~(\ref{G*r}) (\textit{via} Eq.~(\ref{Fg}) with $D_k$ replacing
$g_k$) to the data describing the bead's transient response to the flipping
traps (in which the high-frequency response is limited by the performance of the
SLM).

\begin{figure}
	\centering
		{\includegraphics[width=10cm,height=10cm,keepaspectratio]{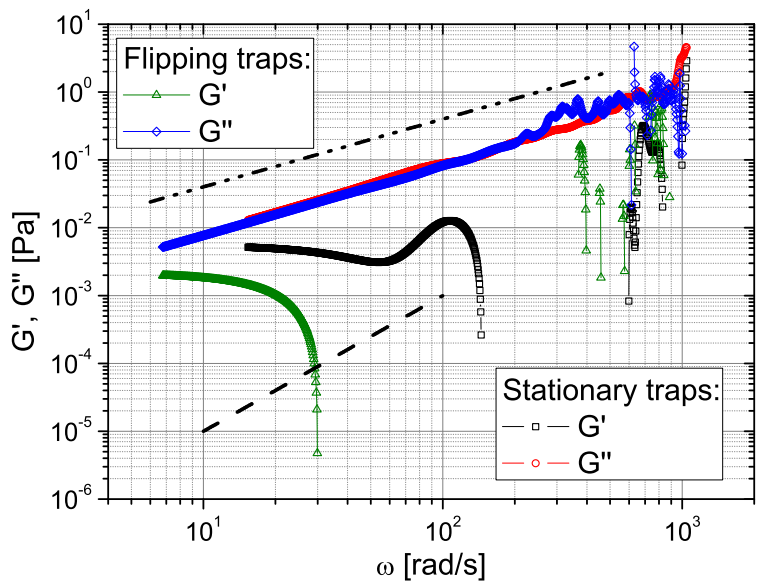}}
	\caption{\label{ModuliW} Storage ($G'$) and loss ($G''$) moduli of water
\textit{vs}.~frequency, analysed using both Eq.~(\ref{G*NPAF})
(high frequencies) and Eq.~(\ref{G*r}) (low frequencies) applied directly to the
experimental data presented in Fig.~\ref{NPAF} and Fig.~\ref{StepN},
respectively. The lines represents the expected limiting behaviour of the moduli
when the material reaches the terminal region: $G'\propto \omega^2$ and
$G''\propto \omega$.}	
\end{figure}

\begin{figure}
	\centering
		{\includegraphics[width=10cm,height=10cm,keepaspectratio]{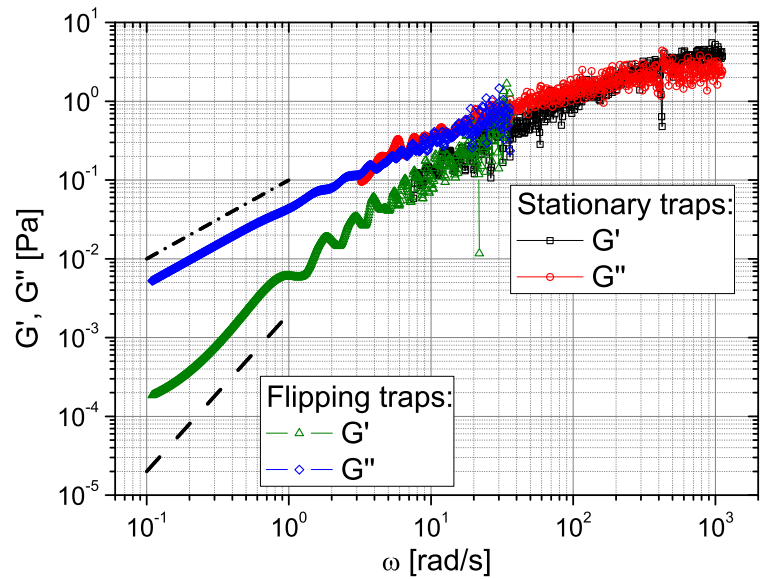}}
	\caption{\label{Moduli} Storage ($G'$) and loss ($G''$) moduli
\textit{vs.} frequency of a solution of 1\% w/w of PAM in water measured by
means of both Eq.~(\ref{G*NPAF}) (high frequencies) and Eq.~(\ref{G*r}) (low
frequencies) applied directly to the experimental data presented in Fig.
\ref{NPAF} and Fig. \ref{StepN}, respectively. The lines represents the expected
limiting behaviour of the moduli when the material reaches the terminal region:
$G'\propto \omega^2$ and $G''\propto \omega$.}	
\end{figure}

Typical results for both Newtonian and non-Newtonian fluids
are shown in Figures~\ref{ModuliW} and \ref{Moduli}, respectively.
In both the cases, it is evident that, although there is some noise in the
frequency domain which comes from genuine experimental noise in the
time-domain data, there is a clear overlapping region of agreement between the
two methods which makes the whole procedure self-consistent. Moreover, it
confirms the ease with which the low-frequency material response can be
explored right down to the terminal region (where $G' \propto \omega^{2}$ and
$G'' \propto \omega$). This is the current limitation for microrheological
measurements.

\section{Conclusions}

In summary, we have presented a self-consistent and simple experimental procedure, coupled with a data analysis methods, for determining the wide-band viscoelastic properties of complex fluids using optical tweezers. This method extends the range of the frequencies previously available to optical tweezers measurements. In fact,the accessible frequency range is limited only by the experiment length and by the maximum data acquisition speed (10s of MHz for a quadrant photo-diode). This allows access to the material's terminal region enabling micro-rheological measurements to be performed on complex fluids with very long relaxation times, such as those exhibiting soft glassy rheology \cite{SollichReview}. The method provides a simple yet concrete basis on which future viscoelastic measurements may be made on both biological and non-biological systems using optical tweezers.
% edited this.

\section{ACKNOWLEDGEMENTS}

The project was funded by the BBSRC, EPSRC DTC and by BT Grant "Listening to the microworld``.

\clearpage 

\section*{References}
\bibliographystyle{unsrt}
%\bibliography{refWBMR.bib}
\bibliography{refWBMR}
% \begin{thebibliography}{10}
% \end{thebibliography}

\end{document}